\documentclass[preprint,12pt]{elsarticle}
\usepackage{graphicx}
\usepackage{amssymb}
\journal{Physics Letters B}
\begin{document}

\begin{frontmatter}

\title{On the RS2 realization of unparticles\tnoteref{preprint}}
\tnotetext[preprint]{preprint LA-UR-08-3550}

\author{Alexander Friedland, Maurizio Giannotti, Michael Graesser}

\address{Theoretical Division, MS B0285, Los Alamos National Laboratory,
Los Alamos, NM 87544, USA}

\begin{abstract}

%  The unparticle physics scenario, in which the fields of the Standard
%  Model (SM) couple to a hidden scale-invariant sector, has attracted
%  a considerable amount of attention.
  To facilitate the study of the unparticle scenario it is very
  desirable to have a treatable model realizing it in four dimensions.
  Motivated by the general idea of the AdS/CFT correspondence, we
  consider a simple construction: the Randall-Sundrum 2 (single brane)
  setup with the Standard Model fields on the brane and a massive
  vector field in the warped bulk.  We show that in this model the
  known properties of vector unparticles -- the nontrivial phase of
  the CFT propagator, the necessity and dominance of contact
  interactions, the unitarity constraint on the conformal dimension of
  the operator, and the tensor structure dictated by conformal
  symmetry -- follow by simple inspection of the brane-to-brane
  propagator.  The phase has a physical interpretation as controlling
  the rate of escape of unparticles into the extra dimension.
  Requiring the correct sign for the imaginary part of the
  longitudinal polarization of the propagator, we obtain the unitarity
  condition $m_5^2\ge0$, which, unlike in the scalar case, is
  unchanged from flat space. This condition results in the unitarity
  bound $d_V\ge3$, or, more generally, $d_V\ge D-1$ for a vector
  unparticle in $D$-dimensional space. It is instructive to consider
  the RS 2 propagator in (Euclidean) position space: at large
  distances it behaves as a \emph{pure} CFT propagator, while at short
  distances it turns into the 5d flat space propagator. The latter is
  softer than the former, thus regulating the would-be divergences of
  the spectral integral and turning the ``contact'' terms seen at low
  energies into finite-range interactions. Upon Fourier transforming
  to momentum space, one finds that at low momenta the CFT piece is
  \emph{subdominant} to the ``contact'' interactions.

\end{abstract}

%\begin{keyword}



%\end{keyword}

\end{frontmatter}


\section{Introduction and motivation}

In anticipation of the LHC, one would like to explore the broadest
range of possible scenarios. Motivated by this, Georgi considered
\cite{GeorgiI,GeorgiII} a class of models with a hidden conformal
field theory (CFT) sector, which couples to the Standard Model (SM) fields
in the ultraviolet (UV). At low energies, this coupling gives rise to
effective interactions between the SM and the CFT.
As an example, a vector current in the SM could be coupled to a vector
operator ${\cal O}_\mu$ in the CFT via
\begin{equation}
  \frac{c_0}{\Lambda^{d-1}} j^{\mu}_{SM} {\cal O}_{\mu},
  \label{unparticle-coupling}
\end{equation}
where $c_0$ is a dimensionless constant and $d$ is the conformal
dimension of ${\cal O}_\mu$, not necessarily an integer.

Models of this type have rather unusual experimental signatures, in
that they do not predict a discrete set of new particles, as in
commonly considered scenarios of TeV-scale extra dimensions or
supersymmetry. Indeed, CFTs do not even have ``in'' or ``out''
states. Rather, the new physics comes in the form of something less
intuitive, dubbed ``unparticle stuff'' by Georgi
\cite{GeorgiI,GeorgiKats}.

How can one think about unparticles?  Georgi originally envisioned a
CFT sector composed of a nonabelian gauge group with a set of new
fermion fields. The coupling constants in this sector were assumed to
flow into a nontrivial infrared (IR) fixed point, {\it a la}
Banks-Zaks \cite{BanksZaks}, at some scale $\Lambda_{trans}$, which is
above the energies of the experiments, but below the mass scale $M$ of
the messenger fields coupling the sector to the SM. This theory shares
many features with QCD \cite{Neubert}. As the hidden sector
``partons'' are produced in a collision of SM particles, they undergo
a QCD-like showering process, in which particles eventually hadronize
if the theory confines at some scale in the far infrared (IR)
\cite{hiddenvalley}. In the limit of exact conformal invariance in the
IR, the showering process never stops, hence the theory does not have
non-interacting ``out'' states.

Many properties of unparticles can be inferred from conformal
invariance alone, without going into the details of the complicated
dynamics.  Using such arguments, unparticles were shown to possess
several curious properties. When produced in the final state, they
behave as a noninteger number of massless particles \cite{GeorgiI}.
When they mediate interactions between SM particles, the corresponding
propagator has a nontrivial phase \cite{GeorgiII}.  Both of these
properties follow from the spectral representation of the ``unparticle
propagator'', which, as argued in \cite{GeorgiII}, by scale invariance
has to have the form\footnote{For simplicity, we consider the scalar case
  here. The vector case, originally considered in
  \protect{\cite{GeorgiII}}, involves important subtleties, as we
  shall see later.}
\begin{eqnarray}
  \label{eq:spectralintegral}
  \langle {\cal O}(p) {\cal O}(-p) \rangle \propto
  \int_0^\infty dM^2  \frac{(M^2)^{d-2}}{p^2-M^2+i\epsilon}
  = \frac{\pi}{\sin d\pi}(p^2)^{d-2} e^{-i(d-2)\pi}
.
\end{eqnarray}
The integral in this equation converges only when $1<d<2$.

Several other crucial features of unparticles have been pointed out
more recently by Grinstein, Intriligator, and Rothstein (GIR)
\cite{GIR}. Using a combination of arguments based on conformal
symmetry and on the properties of the Banks-Zaks model (which is at
weak 't Hooft coupling), GIR found that:
\begin{itemize}
\item The value of $d$ is limited \emph{from below} by unitarity
  \cite{Mack}. (In the context of unparticle physics, this was also
  noted in \cite{Nakayama2007}.) In particular, for primary, gauge
  invariant vector CFT operators, one must have $d_V\ge3$.  Thus, for
  such operators the interval $1<d<2$ on which the integral in
  Eq.~(\ref{eq:spectralintegral}) converges is completely excluded by
  unitarity.
\item The value of $d$ is not limited \emph{from above}.  For
  $d\ge2$, the unparticle scenario must additionally contain contact
  interactions between the SM fields. These contact interactions are
  \emph{necessary} to cure the divergence in the spectral integral.
\item These contact interactions are very important
  phenomenologically, as they \emph{dominate} over the unparticles in 
  SM-SM scattering processes.

\item The tensor structure of the propagator is fixed by the conformal
  group \cite{ADSCFT_Vector}. The propagator in general (for $d_V\ne3$)
  is not transverse, as that would be incompatible with conformal
  symmetry. This affects the rates
  for certain processes involving unparticles.
\end{itemize}

The picture of ``unparticle stuff'' as an ``infinite shower'' provides
important physical intuition, but the resulting complicated set of
QCD-like diagrams is not trivial to deal with, especially at strong
coupling. In particular, the properties laid out above look rather
mysterious in this framework. Does a noninteger number of massless
particles represent something physical in the infinite shower? Since
the presence of an imaginary part in a propagator usually indicates
some intermediate states go on-shell, what goes on-shell in the CFT
sector, which does not have discrete states of definite mass?  Can the
unitarity bound on $d$ and the dominance of the contact terms be
understood in simple, intuitive terms? In fact, we note
that it took some time to realize the crucial features of unparticles
given by GIR.

One may naturally wonder if there exists a different way of thinking
about unparticles that is (i) more intuitive and (ii) more treatable
than a Yang-Mills hidden sector at strong coupling. A natural
candidate to examine are models based on warped extra 
dimensions. The possibility of describing features of the CFT sector
in this framework is strongly suggested by the celebrated AdS/CFT
correspondence \cite{ADSCFT,ADSCFT_Witten}.

The idea has been discussed by several authors, to a varying degree of
detail. Already in \cite{GeorgiI} it is mentioned that scenarios with
infinite extra dimensions, as introduced by Randall and Sundrum
\cite{RSII},  can have
``unparticle-like behavior'' (without elaboration). Ref.
\cite{stephanov} proposes to realize a scenario of deconstructed
unparticles in the two-brane setup. Similar ideas are also
discussed in \cite{Lee}. 
Finally, a recent work by Cacciapaglia,
Marandella, and Terning (CMT) \cite{Terning2008}, containing the most
detailed analysis of the problem to date, arrives at several important
results, as described later.

There seems to be no consensus, however, on whether such
``holographic'' ({\it i.e.}, based on the AdS/CFT correspondence)
constructions should be regarded as genuine models of unparticle
physics.  For example, Ref.~\cite{Neubert}, which appeared after
\cite{stephanov}, remarks that ``to date no explicit model has been
constructed that would exhibit unparticle behavior''.
Refs.~\cite{GeorgiColloquium,GeorgiKats} also do not use AdS
constructions to model unparticles. On the other hand,
Refs.~\cite{Lee,Terning2008} do refer to their respective AdS-based
constructions as unparticle models. Also, Ref.~\cite{softwall} notes
that the soft-wall models in the Randall-Sundrum \cite{RSII} limit
should describe unparticle physics.
Note that the issue does not reduce to questioning
the validity of the AdS/CFT correspondence: a model of unparticles should
incorporate not only the scale-invariant sector, but also the breaking
of scale invariance at the dimensional transmutation scale and
the coupling of the UV theory to the SM \cite{GeorgiKats}.

What basic checks can one perform?  As a true model, an AdS-based
construction must reproduce \emph{all} of the known properties of
unparticles. As an approximate description, on the other hand, it
could reproduce some of the properties, but fail on others ({\it a la}
AdS/QCD \cite{undelivered}).
%Let us see if this is the case. 
Examining the AdS/unparticle literature, one does not get a clear
confirmation that all unparticle properties are present.  While, as
mentioned, the contact terms are seen on the AdS side (as shown by
CMT), the unitarity bounds on the operator dimensions are not
seen. Moreover, some discussions seem to imply an upper bound on $d$,
in contrast to GIR. Even the phase of the propagator, which
uniquely follows from scale invariance \cite{GeorgiII}, does not
explicitly appear. 
%in the AdS models of unparticles. 
Going beyond the unparticle literature, one can find many important
results in the earlier studies of the Randall-Sundrum (RS) models and
their connection to the AdS/CFT correspondence,
{\it e.g.},
\cite{RS2CFToriginal,Gubser:1999vj,GiddingsKatzRandall,ArkaniHamed:2000ds,Dubovsky,PerezVictoria2001}.
In particular, properties such as the contact terms
\cite{PerezVictoria2001} and the imaginary part of the propagator
\cite{Dubovsky,PerezVictoria2001} were noted.  Other questions,
however, {\it e.g.}, the unitarity properties of the RS
models do not seem to have been investigated.

In what follows, we consider a very simple scenario: the
Randall-Sundrum 2 (RS 2) model where a single positive tension
Minkowski brane is sandwiched between two anti-deSitter 5d regions.
The SM fields are localized on the brane and a single massive vector
field can propagate in the bulk.
%As we will explicitly see, the bulk field does not give just a
%pure CFT on the brane.
%This is in accord
%with the observations of \cite{GeorgiKats} that the existence of the
%UV CFT breaking scale is physically very important.
When considered at scales much below the AdS curvature $\kappa$, this
model will be seen to satisfy all of the properties of unparticles
listed above: the phase of the CFT propagator, the contact terms and
cancellations, the unitarity bounds, and the tensor structure.
Moreover, we will see that the model resolves the
singularities of the ``contact terms'': they correspond to finite
range interactions and disappear above the scale $\kappa$.  Our 
analysis also almost trivially leads to the unitarity bounds on the
conformal dimension of CFT vector operators in $D$ spacetime
dimensions. These and other results are discussed after all of the
properties of unparticles are confirmed.

In the interests of clarity, this paper focuses on conceptual points.
Detailed derivations will be given in the
companion paper \cite{longpaper}.

\section{Properties of unparticles from the RS 2 realization}
\label{sect:properties}

% We seek the simplest model that accomplishes all this. Such a model is
% a single vector field in the bulk with SM fields on the brane. The
% bulk vector then mediates interactions between the Standard Model
% fields. Although similar setups have been extensively studied in the
% literature, many of the key features relevant to our goal do not seem
% to be clearly discussed.

As mentioned, we consider a model with
a single vector field in the RS 2 bulk and SM fields on the brane.
The physical meaning of the unparticle propagator is simple: it is
the brane-to-brane propagator of the bulk vector.

The propagator is the Green's function of the following 
equation \cite{longpaper} ($a=e^{-\kappa|x_5|}$ denotes the warp
factor in the metric, $g_{MN}=diag(a^2\eta_{\mu\nu}, -1)$): 
\begin{equation}\label{eq:master_greensfunction}
  -(p^2 \eta_{\mu\nu}- p_\mu p_\nu) A^{\mu} -
  \partial_5\left(a^2\left[\eta_{\mu\nu}
      -\frac{p_\mu p_\nu}{p^2-m^2a^2}\right]\partial_5 A^{\mu} \right)
  + m^2_5 a^2 A_\nu = 0 ~.
\end{equation}
 Two crucial
physical ingredients are as follows.
\begin{itemize}
\item First, the field has a bulk mass, $m_5$. In AdS/CFT,
  this mass is known to control the conformal dimension of the
  resulting vector CFT operator \cite{ADSCFT_Witten}.
With $m_5\ne0$, the vector field
  has four degrees of freedom. From the 4-dimensional point of view,
  it has three transverse polarizations ($\epsilon_\mu^{(\alpha)} p^\mu=0$)
  and one longitudinal ($\epsilon^{\mu}\parallel p^\mu$). The
  longitudinal component is often omitted in the literature. It is
  important to keep it: first, it is essential for obtaining the
  correct tensor structure that respects conformal symmetry; second,
  it is crucial for our unitarity arguments.
\item Second, in deriving the propagator (Green's function), we need
  to fix the boundary conditions away from the brane. Physically,
  collisions of the
  SM particles on the brane result in outgoing waves of the
  bulk vector. The corresponding boundary condition is known as the
  Hartle-Hawking or radiative \cite{GiddingsKatzRandall,Dubovsky}. It
  is also obtained if one rotates the modified Bessel function
  solutions giving finite action \cite{ADSCFT_Witten} from Euclidean to Minkowski space.
\end{itemize}
With these ingredients, one obtains the vector boson propagator
\begin{eqnarray}
\label{corr_decompose}
   \Delta_{\rho\sigma}(p^2) &=&
   \left(-\eta_{\rho\sigma} +\frac{p_\rho p_\sigma}{p^2}\right)\Delta_T(p^2)
   -\frac{p_\rho p_\sigma}{p^2}\Delta_L(p^2),\\
\Delta_T(p^2)  &=&  \frac{1}{2}\left[ p
\frac{H_{\nu-1}^{(1)}(p/\kappa)}{H_{\nu}^{(1)}(p/\kappa)}
-\kappa(\nu -1) \right] ^{-1},
\label{transverseP} \\
\Delta_L (p^2)  &=& \frac{1}{2 m^2_5} \left[  p
\frac{H_{\nu-1}^{(1)}(p/\kappa)}{H_{\nu}^{(1)}(p/\kappa)}
-\kappa(\nu +1) \right]. \label{longitudinalP}
\end{eqnarray}
 where  $p\equiv\sqrt{p^2}$, $H_{\nu}^{(1)}(x)$ is the Hankel function
 of the first kind and 
%$\nu$ is defined in Eq.~(\ref{eq:nudefine}) above.
\begin{equation}
  \label{eq:nudefine}
  \nu =\sqrt{1+ \frac{m^2_5}{\kappa^2}}.
\end{equation}
This propagator is easily obtained by taking the outgoing wave
solutions of Eq.~(\ref{eq:master_greensfunction}) in the
bulk and imposing matching conditions on the brane ({\it cf., e.g.,}
\cite{Dubovsky}). For the transverse polarizations, the bulk solution
is $c(p)e^{\kappa |x_5|} H_\nu^{(1)}(e^{\kappa |x_5|}p/\kappa)$, and
$c(p)$ is fixed from the condition
$\partial_5\Delta_T(p^2,+\epsilon)-\partial_5\Delta_T(p^2,-\epsilon)=1$.
For the longitudinal mode, a similar bulk solution exists for the
combination $g(p^2,x_5)\equiv
a^4\partial_5\Delta_L(p^2,x_5)/(p^2-m_5^2a^2)$, with the matching
condition $g(p^2,+\epsilon)-g(p^2,-\epsilon)=-1/m_5^2$. See
\cite{longpaper} for details.

The properties of unparticles due to Georgi and GIR listed in the introduction
follow from this propagator by simple inspection. 

\emph{Contact terms}: Obviously,
Eqs.~(\ref{corr_decompose},\ref{transverseP},\ref{longitudinalP}) do
not describe a pure CFT propagator. In a pure CFT, one expects $x^{-2d}$
everywhere, or in momentum space $p^{2d-4}$.
%[$p^{2d-n}$ in $n$-dim spacetime, cf. Georgi-Kats].
The RS 2 propagator, in the $p \rightarrow 0$ limit, looks instead
like a contact interaction, $\Delta^{(L)}(0) = -\kappa(1+\nu)/2m^2_5$,
$\Delta^{(T)}(0) = -1/2\kappa(\nu-1)$. Expanding the propagator in
series for small $p/k$, we get for the longitudinal part
\begin{eqnarray}\label{eq:longitudinal_branetobrane_expand_Mink}
          \Delta^{(L)} (p^2) &\simeq& \frac{\kappa}{2m^2_5}\left[
          -(1+\nu) +
          \frac{(p/\kappa)^2}{2(\nu-1)} +
          \frac{(p/\kappa)^4}{8(\nu-1)^2(\nu-2)} \right.\nonumber\\
       &+&
          \frac{(p/\kappa)^6}{16(\nu-1)^3(\nu-2)(\nu-3)} +
       \frac{(5\nu-11)(p/\kappa)^8}{128(\nu-1)^4(\nu-2)^2(\nu-3)(\nu-4)}+ \cdots
\nonumber\\
       &+&\left.\frac{2 \pi}{\Gamma(\nu)^2}(i - \cot\pi\nu)\left(\frac{p}{2 \kappa} \right)^{2\nu}\left[1+\cdots \right]\right]
     .
\end{eqnarray}
The transverse part has a very similar expansion, in the denominator.

The terms in the first and second lines have the structure of contact
terms: const, $p^2$, $p^4$, etc.  (Corresponding to the Fourier
transform of $\delta(x)$, $\partial^2 \delta(x)$, $\partial^4
\delta(x)$, etc.) Assuming for the moment that $\nu > 1$, we see that
the nonanalytic terms in the last line are subdominant. As we will see
shortly, $\nu \ge 1$ is indeed required, by unitarity. Hence, the
dominance of the contact terms is established.

Two important caveats must be mentioned at this point: First, the case
$\nu=1$ involves some subtleties that will be discussed in the
companion paper \cite{longpaper}. Here, we will assume $\nu$ is not
too close to 1. Second, although we just referred to the terms in the
first two lines of
Eq.~(\ref{eq:longitudinal_branetobrane_expand_Mink}) as ``contact''
terms, this is not strictly accurate. The whole series describes an
interaction \emph{with a finite range}, as will be discussed later.

\emph{CFT}: We now turn to the last line of
Eq.~(\ref{eq:longitudinal_branetobrane_expand_Mink}). The
leading nonanalytic term in the expansion behaves as $p^{2\nu}$,
exactly as expected for a CFT with
\begin{equation}
  \label{eq:dfromnu}
  d_V=\nu+2.
\end{equation} 
In fact, the following general result is well known 
\cite{OzTerning,BigReview},
\begin{equation}
  \label{eq:dforgeneralD}
  (d-p)(d+p-D)=m_5^2/\kappa^2.
\end{equation}
For the vector field (a 1-form), $p=1$; the spacetime dimension on the
brane is $D=4$, indeed yielding Eq.~(\ref{eq:dfromnu}), with $\nu$ as
in Eq.~(\ref{eq:nudefine}).

To find the leading nonanalytic piece for the transverse mode, we
expand in series, assuming $\nu - 1$ is not too small, as already
mentioned.
\begin{eqnarray}
  \label{eq:transverseCFT}
\Delta_T(p^2) &\simeq&  \frac{1}{2\kappa}\left[
-(\nu -1) + ... +  \frac{2 \pi(i - \cot\pi\nu)}{\Gamma(\nu)^2}
  \left(\frac{p}{2 \kappa}\right)^{2\nu} + ...
     \right]^{-1}  \nonumber\\
     &\simeq&
     \frac{1}{2\kappa}\left[-\frac{1}{\nu -1} + ... - \frac{1}{(\nu -1)^{2}}
     \frac{2 \pi(i - \cot\pi\nu)}{\Gamma(\nu)^2}
     \left(\frac{p}{2 \kappa}\right)^{2\nu} + ...\right].\;\;\;\;
\end{eqnarray}
The RS 2 propagator thus \emph{contains} the CFT part, both in its
transverse and longitudinal components, at a subleading
order in $p/\kappa$.

\emph{The phase}: Notice that while the analytical terms in the
expansion of the propagator are purely real, the CFT piece does have
both real and imaginary parts. Notice that
$i-\cot\pi\nu=-\exp(-i\pi\nu)/\sin\pi\nu$. Since $d=\nu+2$, the
nonanalytic terms, both longitudinal and transverse, have exactly the
phase of the unparticle propagator, Eq.~(\ref{eq:spectralintegral}).

\emph{CFT tensor structure}: Combining the transverse and longitudinal
nonanalytic parts, we see that the correct CFT tensor structure
nontrivially emerges ({\it cf.}  \cite{ADSCFT_Vector}). Explicitly,
combining the leading nonanalytic parts from
Eqs.~(\ref{eq:longitudinal_branetobrane_expand_Mink}) and
(\ref{eq:transverseCFT}) according to Eq.~(\ref{corr_decompose}) and
recalling that $m_5^2=\kappa^2 (\nu^2-1)$ from
Eq.~(\ref{eq:nudefine}), we get
\begin{eqnarray}\label{correcttensor}
    \Delta_{\mu\nu}^{non-analyt}=
  C \frac{\pi e^{-i\pi\nu}}{\sin\pi\nu}
      p^{2\nu}
    \left(-\eta_{\mu\nu} +\frac{2\nu}{\nu+1}\frac{p_\mu p_\nu}{p^2}\right),
 \end{eqnarray}
where the overall constant is $C\equiv [(\nu-1)^2\Gamma(\nu)^22^{2\nu}
\kappa^{2\nu+1}]^{-1}$.

\emph{The divergences}: Note that $\Delta_{\mu\nu}^{non-analyt}$ in
Eq.~(\ref{correcttensor}) becomes singular at integer values of $\nu$
(and hence $d$). On the other hand, the full propagator, being a
combination of well defined Bessel functions, is perfectly well
defined for any $\nu\ge1$. Indeed, physically there is nothing special
about the values of $m_5$ that yield integer $\nu$.  This means the
singularities in $\Delta_{\mu\nu}^{non-analyt}$ must be cancelled by
the corresponding singularities in the analytic terms. Indeed, from
Eq.~(\ref{eq:longitudinal_branetobrane_expand_Mink}), we explicitly
see how this cancellation takes place: the residues of the poles of
the cotangent match the corresponding residues of the poles of the
analytic terms.

Thus, the cancellations between the singularities of the ``CFT
propagator'' and the ``contact terms'' in the RS 2 picture reduce
simply to a well known property of the expansion of the Bessel
functions. One should not be alarmed that individual terms in the
expansions become singular, after all the function is being expanded
around its branch point. We will return to the analytic properties of
the propagator later.

\emph{Unitarity}: The imaginary part of the propagator has a direct
physical meaning in the RS 2 setup, as will be discussed later.
Namely, it tells us about the width of the ``escape into extra
dimensions''. This width clearly has to be nonnegative, not to violate
unitarity. This must hold for any choice of the wavefunctions that the
propagator is sandwiched between. Therefore, it must hold separately
for the transverse and longitudinal modes. The relative sign between
them can only be correct if 
\begin{equation}
  \label{eq:ourunitarity}
  m_5^2>0,
\end{equation}
as follows from inspecting Eqs.~(\ref{transverseP}) and
(\ref{longitudinalP}).  It can be shown that this condition is not
only necessary, but also sufficient for unitarity \cite{longpaper}.
Notice that this seemingly trivial condition differs, for example,
from the scalar case. A scalar field in the AdS$_{D+1}$
background can in fact have a \emph{negative} mass-squared, so long as
$m_5^2/\kappa^2\ge-D^2/4$ \cite{Freedmanetal,ADSCFT_Witten}.  

Substituting $m_5^2$ in the definition of $d$, we
find $d_V\ge3$, exactly the condition derived by Mack
\cite{Mack}. Eq.~(\ref{eq:ourunitarity}) in fact turns out to be more
general, as discussed later.

\section{Discussion}

%The fact that the RS 2 model realizes the unparticle scenario makes it
%possible to address a variety of questions.
We now discuss several issues in unparticle physics, for which
the RS 2 realization provides helpful insights.

\emph{Spectral representation as a sum over KK modes}: Let us return
to the spectral representation, Eq.~(\ref{eq:spectralintegral}).
In the RS 2 model, this equation describes a physical sum over a
continuous spectrum of single-particle states, the Kaluza-Klein (KK)
modes of the bulk field. In this realization, the term ``unparticle
stuff'' \cite{GeorgiI,GeorgiKats} gains a simple physical meaning: it
is a tower of particles with a continuous spectrum and couplings that
scale as a power of mass. The unparticle scenario can then be viewed
as a case \cite{Krasnikov2007} of a more general framework in which
fields have continuously distributed mass \cite{Krasnikov1994}.

Notably, Ref.~\cite{Krasnikov1994} specifically proposed to realize a
scalar field with continuously distributed mass as a scalar living in
a flat five-dimensional spacetime coupling to the SM fields on the
brane.  The connection between flat extra dimensions and unparticles
was also noted in, {\it e.g.}, \cite{Cheung:2007PRD}.

What the RS 2 construction brings to this picture is a way to control
the relative couplings of different KK modes to fields on the brane.
The physics of this is most transparent upon transforming the field
equation for the bulk field into the form of the Schr\"odinger
equation.  This transformation is discussed in the original RS 2 paper
\cite{RSII} and, {\it e.g.}, in \cite{ourprl} (in the context of the
RS 2 scenarios extended with additional compact dimensions
\cite{extraextradim}).  In the Schr\"odinger description, the
low-energy bulk KK modes have to tunnel through a potential $V(s)=f
s^{-2}$ to reach the brane, where $f$ is a function of $m_5/\kappa$.
As can be easily seen \cite{ourprl}, for this particular potential the
wavefunctions on the brane are \emph{power-law}, rather than
exponentially, suppressed, leading to a CFT behavior. The power
is determined by the function $f$, and hence by the value of the bulk
mass $m_5$, yielding Eq.~(\ref{eq:nudefine}).

\emph{The phase}: The presence of the imaginary part in the unparticle
propagator also gains a physical meaning in the RS 2 realization. It
is clear what goes on-shell in the propagator: collisions of the SM
particles on the brane can excite bulk KK modes with the right mass.
The unparticles ``leak out'' into the bulk because of the incomplete
gravitational binding (or, equivalently tunnel under the volcano
potential). The physical discussion of this
escape is given in Refs.  \cite{Dubovsky,RubakovReview}. This escape
corresponds to the unparticle production. Even though in an
experiment, it appears as the production of an integer or noninteger
number of massless four-dimensional particles (``neutrinos'')
\cite{GeorgiI}, in the RS 2 model what is produced is a single KK
plane wave, which is massive from the four-dimensional point of view.
Notice that, once escaped, the bulk particles do not re-interact with
the brane fields, unless the extra dimension is compactified ({\it
  cf.}  \cite{stephanov}).

Notice also that the analytic (``contact'') terms in the expansion of the
RS 2 propagator do not have imaginary parts.  Physically, this is
because they correspond to heavy degrees of freedom, as will be
discussed shortly. These heavy degrees of freedom are exchanged
far off-shell, without exciting asymptotic outgoing plane waves far from
the brane. 

Since the analytic terms dominate at low $p$, the phase of the complete
propagator is rather small. The phase factor multiplying the
nonanalytic piece, $e^{-i\pi(d-2)}$, by itself is not particularly
meaningful: for example, the fact that it becomes real at integer $d$
in no way implies that the unparticles do not escape from the brane in
that case. All that happens is the $\cot\pi\nu$ term in
Eq.~(\ref{eq:longitudinal_branetobrane_expand_Mink}) blows up there;
this divergence is, however, cancelled by the corresponding analytic
term. On the other hand, \emph{the imaginary part} of the nonanalytic
piece is a meaningful quantity, telling us the escape rate
\cite{Dubovsky}.

\emph{In and Out states}: Comparing the CFT and the AdS descriptions,
one may notice the following apparent ``paradox''. The CFT does not
have ``out'' states: what is produced in a SM-SM collision is a ball
of hidden sector quarks and gluons that continues to expand as the shower
develops. In contrast, in the RS 2 setup what is produced is a
wavepacket propagating in the bulk, which corresponds to a notion of
an ``out'' state. Notice, however, that just like one cannot reverse
the development of the shower and use it as the ``in'' state for
producing SM particles, the wavepacket in the RS 2 description, once
emitted into the bulk, cannot be accessed from the brane.

\emph{Unitarity bounds in $D$ dimensions}: Let us now elaborate on the
unitarity condition derived in Sect.~\ref{sect:properties}. The bound
followed from the requirement that the imaginary part of the
longitudinal component of the propagator have the right sign,
corresponding to particles escaping from the brane. Observe that our
argument at no point relied on having exactly four spacetime
dimensions on the brane.  This means that $m_5^2>0$ is in fact a more
general unitarity condition, valid for vector CFT operators in a
$D$-dimensional Minkowski spacetime.  To convert it into a bound on
the dimension of the conformal operator, recall that for general $D$
the relationship between $\nu$ and $m_5^2$ (or directly between $d$
and $m_5^2$) is given in Eq.~(\ref{eq:dforgeneralD}).  Therefore,
$m_5^2\ge0$ gives
\begin{equation}
  \label{eq:unitarityD}
  d_{unit}\ge D-p=D-1.
\end{equation}
This agrees with the results of \cite{Minwalla}. The RS 2 construction
thus gives a remarkably simple derivation of this result.

Two more comments should be made. First, the unitarity bound is known
to apply to gauge invariant, primary operators. Indeed, our
\emph{massive} vector field does not have any gauge degrees of freedom
and is not a derivative of a scalar. Second, the bound on the CFT
vector operator comes from imposing the positivity constraint on the
coefficient of the correlator of the first descendent operator
($\langle\partial^\mu {\mathcal O}_\mu(x)\partial^\nu {\mathcal
  O}_\nu(0)\rangle$) \cite{GIR}.  This is consistent with the fact
that our bound comes from the \emph{longitudinal} part of the
propagator.

\emph{UV limit -- flat space}: Now, let us examine what happens in the
limit $p\gg\kappa$. If we also assume $m_5^2 \gg \kappa$,
the propagator takes the form \cite{longpaper}
\begin{eqnarray}
  \label{eq:flat_proca_brane}
  \Delta_{\mu\nu}^{flat}(p^2) &=&
  \left(-\eta_{\mu\nu} +\frac{p_\mu p_\nu}{p_4^2}\right)
\frac{1}{2}\frac{-i}{\sqrt{p_4^2-m^2_5}}
-\frac{p_\mu p_\nu}{p_4^2}\frac{i}{2m^2_5}\sqrt{p_4^2-m^2_5}  \\
& \equiv & \left(-\eta_{\mu\nu} +\frac{p_\mu p_\nu}{p_4^2}\right) \Delta^{flat}_T(p^2_4)
 -\frac{p_\mu p_\nu}{p_4^2} \Delta^{flat}_L(p^2_4)
\end{eqnarray}
This is nothing but the propagator of the massive field in
flat 5d space. Indeed, the tensor structure of the standard 5d vector
propagator, $(-\eta_{MN} + P_M P_N/m^2_5)/(P^2-m^2_5 +i \epsilon)$,
can be decomposed into transverse and longitudinal parts according to
$-\eta_{\mu\nu} + p_\mu p_\nu/m^2_5= [-\eta_{\mu\nu} + p_\mu
p_\nu/p^2_4] + (p_\mu p_\nu/p_4^2) (p_4^2-m^2_5)/m^2_5$. Upon
integrating over $p_5$, we find
Eq.~(\ref{eq:flat_proca_brane}).

The propagator in Eq.~(\ref{eq:flat_proca_brane}) is seen to be
\emph{purely imaginary}. Physically, this means that in flat space the
KK modes can freely escape into extra dimensions. At low momenta,
$p\ll\kappa$, gravity provides (incomplete) confinement to the brane;
for $p\gg\kappa$, this confinement is negligible.

The RS 2 model thus completes the unparticle scenario in the UV with
5d flat space. This is a different completion from the asymptotically
free 4d Yang-Mills theory envisioned by Georgi. Nevertheless, it
adequately regulates the theory, as we will see next\footnote{It must
  be mentioned that our treatment here is to the leading order. To
  better define the theory in the UV, one may want to replace the mass
  term by the Higgs mechanism and also to deconstruct the
  5-dimensional theory (latticize the $x_5$ coordinate) on scales
  shorter than the AdS curvature.}.

\emph{More on flat space: unitarity}.  Consider now the limit $p \gg
m_5$ of Eq.~(\ref{eq:flat_proca_brane}). The transverse propagator
looks ``unparticle-like'', with dimension 3/2, which is just the
engineering dimension of the vector field in 5d. It is important that
this does not violate the unitarity bound on vector operators in CFT:
the bound only applies to gauge invariant operators and in this limit
the longitudinal polarization becomes a gauge degree of freedom. (The
longitudinal propagator is seen to blow up.) For finite $m_5$, all
four polarizations on the brane are physical, but the theory in the
flat space limit is not conformal. This example illustrates some sense
in which it is possible to consider phenomenological signatures of
``vector unparticles'' with dimensions below the $d_V\ge3$ bound: a
vector field that can propagate in flat extra dimensions could be
interpreted experimentally as a vector unparticle with the ``wrong''
dimension.  A less trivial example of gauge-variant vector unparticles
could be provided by models with additional warped compact dimensions
\cite{extraextradim}, in which the ``photon-unparticle'' decay (photon
escape into extra dimensions) is possible \cite{RubakovReview,ourprl}.

\emph{Still more on flat space: no contact terms}. One more important
observation about the flat space limit needs to be made: the
propagator in this limit has a cut, but \emph{no contact terms}.
Comparing with the $p\ll\kappa$ expansion,
Eq.~(\ref{eq:longitudinal_branetobrane_expand_Mink}), we see that
``the contact terms'' seen at low energies disappear above the scale
$\sim\kappa$. This implies that these terms are not fundamental
point-like interactions, but in fact have finite range, $\gtrsim
\kappa^{-1}$. This also suggests that the flat space UV regime of the
RS 2 model softens and regularizes what would otherwise be a divergent
behavior of the pure CFT.  To understand the implications of this
better, let us consider in turn the spectral representation of the RS
2 propagator and its behavior in position space.

\emph{Spectral representation: regularization by flat space.} In
light of our observations about the flat space limit, let us reexamine
the spectral representation of the unparticle propagator. As already
mentioned, the integral in Eq.~(\ref{eq:spectralintegral}) in the RS 2
model becomes a sum over the KK states. The couplings of the states
with masses below the curvature scale grow as $(M^2)^{d-2}$, but the
growth is halted at the scale $\kappa$. Those states instead behave as
in the 5-dimensional flat space, meaning that they couple to the brane
with equal strengths. Schematically, in the RS 2 model one can write
(omitting factors),
\begin{eqnarray}
  \label{eq:spectralintegral2}
  \langle {\cal O}(p) {\cal O}(-p) \rangle \sim
  \int_0^{\kappa^2} dM^2  \frac{(M^2)^{d-2}}{p^2-M^2+i\epsilon}+
  \int_{\kappa^2}^\infty \frac{dM^2}{M}  \frac{(\kappa^2)^{d-3/2}}{p^2-M^2+i\epsilon}.
\end{eqnarray}
In the second integral, the measure of integration $dM^2/M$ comes from
$dp_5$.

The second integral converges, yielding $\sim\kappa^{d-2}$ for
$\kappa\gg p$. For $d<2$, this becomes infinitesimally small. The
upper limit in the first integral can then be extended to infinity,
recovering the spectral representation of \cite{GeorgiII},
Eq.~(\ref{eq:spectralintegral}). Physically, for $1<d<2$ the
interactions involving exchange of momentum $p$ is dominated by modes
with masses not much greater than $\sim p$ and the contribution of the
UV tail is negligible. For $d\ge2$, on the other hand, the
contributions of the heavy states ($M\gg p$) dominate the integral.
The answer in that case is sensitive to the physics in the UV by
construction \cite{Terning2008} and diverges as the upper integration
limit is taken to infinity.

It is physically clear that the interactions dominated by
short-distance modes has to look like contact terms at low energies
($p\ll\kappa$). Indeed, it is easy to show that in the limit
$\kappa\rightarrow\infty$ the divergent part of the spectral function
is localized to contact terms.  For this, we observe that
differentiating and integrating back with respect to $p^2$ drops the
$\delta(x)$ counterterm (constant in momentum space). Differentiating
Eq.~(\ref{eq:spectralintegral}) with respect to $p^2$ yields an
integral that converges for $d<3$. Thus, the divergence of the
integral for $2<d<3$ is in the additive constant. Similarly,
differentiating twice extends the interval of convergence to $d<4$.
Notice that the improved convergence of the spectral integral upon the
subtraction of local terms was shown earlier by CMT
\cite{Terning2008}.

It is important to stress that the dominant part of the spectral
integral becomes local and divergent only in the limit of
$\kappa\rightarrow\infty$. For finite $\kappa$, everything is regular,
as we have seen explicitly in studying the propagator,
Eq.~(\ref{corr_decompose},\ref{transverseP},\ref{longitudinalP}). The
RS 2 model does not require fundamental contact counterterms.

Generalizing the convergence argument above, we can write
\begin{eqnarray}
  \label{eq:spectralintegralcontact}
  \int_0^{\kappa^2} dM^2  \frac{(M^2)^{d-2}}{p^2-M^2+i\epsilon}
&=&
  \frac{ \pi}{\sin d\pi}(p^2)^{d-2} e^{-i(d-2)\pi} + a_0 + a_1 p^2 +
  ...
\nonumber\\
 &+& a_{[d-2]} (p^2)^{[d-2]} + \cdots,
\end{eqnarray}
where $[d]$ denotes the greatest integer less than $d$ and the
coefficients $a_n$ diverge as $\kappa^{2([d]-2-n)}$ with the cut-off
of the integral. The low-energy  expansion of the RS 2 propagator,
Eq.~(\ref{eq:longitudinal_branetobrane_expand_Mink}), has exactly this
form.

The argument that for $d>2$ the spectral integral is dominated by
short-distance physics is quite general and applies to any realization
of the unparticle scenario, not just RS 2. All that is necessary is
that the integral be \emph{somehow} regularized above the
transmutation scale in the UV. 
%at the scale at which the theory flows into the fixed point.  
In fact, the argument
generalizes to a broader set of models with a continuous spectrum of
excitations, not only CFTs.  
% So long as the spectral integral is
% dominated by the UV contributions, the contact terms dominate the
% SM-SM scattering process. In this sense, the contact terms are a more
% robust feature than the unitarity bounds, which, as we saw, could be
% avoided by relaxing the conditions of gauge invariance, or strict
% conformal symmetry.

\begin{figure}
  \centering
  \includegraphics[angle=0,width=0.90\textwidth]{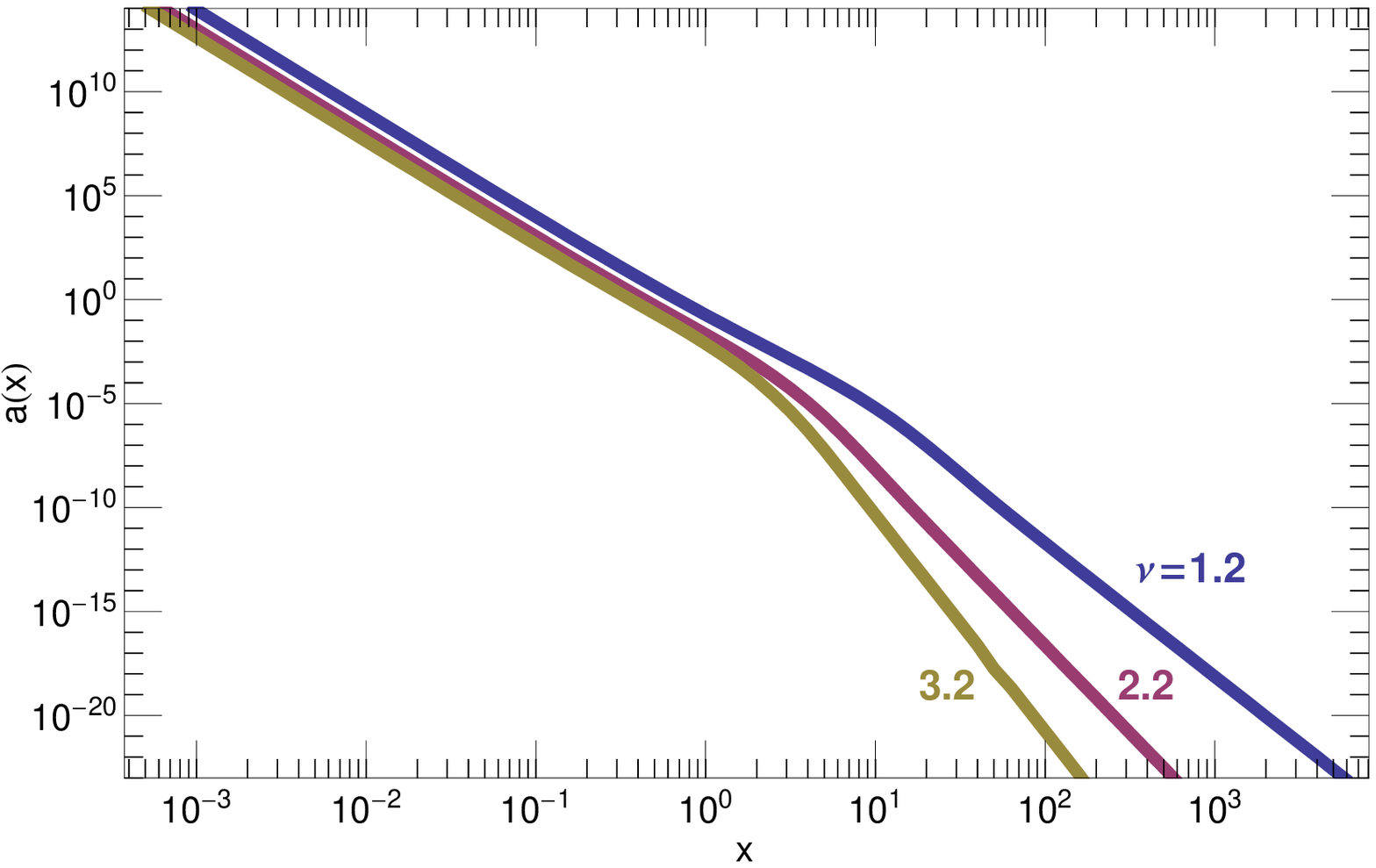}
  \includegraphics[angle=0,width=0.90\textwidth]{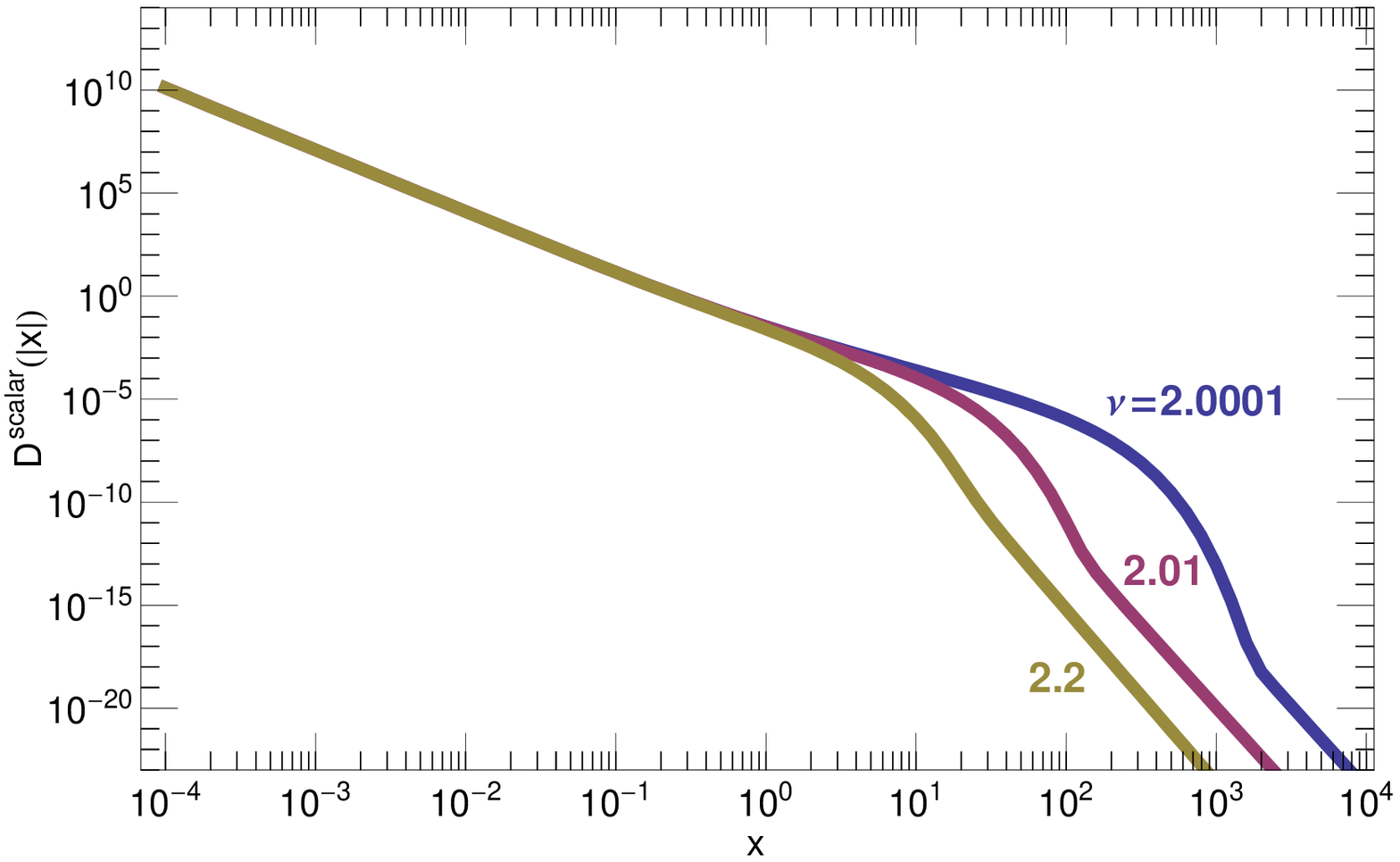}
  \caption{The Euclidean Green's function of the vector (top) and
    scalar (bottom) fields in position space. For the vector field, we
    plot the function $a(x)$, defined by $ D_{ij} (x)= a(x)
    \delta_{ij} + b(x) x_{i} x_{j}/x^2$. For simplicity, the AdS
    curvature $\kappa$ was set to 1, {\it i.e.}, the distance $x$ is
    in units of $\kappa^{-1}$. Three different values of $\nu$ are
    considered, as labeled in the plots. In both cases, the functions
    exhibit two power law regimes: 5d flat space at short distances
    ($x < 1$) and the CFT at long distances ($x \gg 1$). The 
    transition behavior is seen to differ for the two: the scalar has
    a pronounced regime where the localized mode dominates.}
  \label{fig:Greens_position}
\end{figure}

\emph{Propagator in position space.}  An important insight into 
the physics of the RS 2 model comes from considering the behavior of the
propagator in (Euclidean) position space.  Fourier transforming the
vector propagator, we obtain an expression of the form $ D_{ij} (x)=
a(x) \delta_{ij} + b(x) x_{i} x_{j}/x^2$. In
Fig.~\ref{fig:Greens_position} for illustration we plot $a(x)$ for
several values of $\nu$ (top panel). The corresponding plot for $b(x)$
is qualitatively similar \cite{longpaper}. We also show, for
comparison, the position space correlator for a scalar
field (bottom panel).

We see that, in both cases, the propagator at short distances takes
the form expected in five-dimensional flat space. Moreover,
importantly, at long distances it goes into \emph{the pure} CFT
regime. The two regimes appear as power laws (straight lines) in the
Figure. The slopes of the lines (the conformal dimensions) at large
distances depend on $\nu$ as in  Eq.~(\ref{eq:nudefine}),
while at short distances all lines have the same slope, characteristic
of the flat 5d space.  Importantly, the curves become less steep at
short distances: the flat space completion \emph{softens} the
propagator, as already noted.

The transition between CFT and flat space (``transmutation'') for
generic $\nu$ is seen to occur at the scale of the AdS curvature. When
$\nu$ is close to 1 for the vector or 2 for the scalar, {\it i.e.}, when
$m_5\ll\kappa$, the transition window becomes extended.
The
details of the transition are seen to differ for the vector and scalar
cases.  For the scalar, in the transition window 
%there is a range of $x$, in which 
the interaction is dominated by a mode bound to the brane, i.e., the
interaction becomes four-dimensional \cite{Dubovsky}. The vector
behaves differently. These details are beyond the scope of this paper
and will be discussed in \cite{longpaper}.

Here, we wish to discuss two qualitatively important -- if seemingly
paradoxical -- features seen numerically in
Fig.~\ref{fig:Greens_position}: (i) The pure CFT form of the
propagator at long distances [$\propto(x^2)^{-2-\nu}$] suggests it is
a Fourier transform of $p^{2\nu}$. Yet, we have seen that in the
momentum space expansion ({\it cf.}
Eqs.~(\ref{eq:longitudinal_branetobrane_expand_Mink},
\ref{eq:transverseCFT})) the $p^{2\nu}$ term is \emph{subdominant}.
(ii) The leading terms in the low-momentum expansion
($p^0$, $p^2$, ...) have the form of contact interactions.  Yet,
contrary to some discussions in the literature, no contact
interactions are seen in position space.

To understand point (i), consider the following mathematical property
of Fourier transforms \cite{Migdal}. If a function is analytic
everywhere on the real axis, its high frequency Fourier modes are
suppressed \emph{exponentially}. The exponential factor is determined
by the distance from the closest singularity to the real axis. When
the function has a point of nonanalyticity on the real axis, its high
frequency Fourier modes are only power suppressed. Simply put,
``sharp features'' of the function (discontinuities, cusps, etc) carry
the high frequency signal.

In light of this, consider the RS 2 propagator. Schematically, it is
the sum of two pieces, one of which ($p^{2\nu}$) has a singularity on
the real axis (at $p^2=0$). This singularity dominates the Fourier
transform at large ``frequencies'', {\it i.e.}, large position space
distances. That $p^{2\nu}$ occurs at a subleading order in the
small-$p^2$ expansion does not change this conclusion.

Now consider point (ii). Observe that to find the small-distance
behavior of the correlator we need its large-momentum behavior.  Yet,
the latter is not obvious from the first several terms of the
expansion around $p^2=0$. Indeed, a series around the origin describes
a function only inside a circle up to the nearest singularity. A
simple illustration is provided by a massive scalar field in four
dimensions: the interaction has a finite range, $\sim m^{-1}$, even
though transforming the small-$p^2$ expansion of $[p^2+m^2]^{-1}$
term-by-term one would get a series of contact terms
$\partial^{2n}\delta^{(4)}(x)/m^{2n+2}$. The ``nonperturbative
effect'' $e^{-m x}$ is missed in this series.  In our case, the RS 2
propagator is being expanded around its branch point and hence the
radius of convergence of the series is zero. The large-momentum
behavior of the propagator is instead given in
Eq.~(\ref{eq:flat_proca_brane}), from which we explicitly get no
contact terms.

The position-space picture gives a simple prescription for describing
the unparticle propagator in a generic realization. Take the
propagator to be of the form $(g_{\mu\nu}-2 x_\mu x_\nu/x^2)/x^{2d}$
\cite{GIR} at large distances. This respects the conformal symmetry.
Now, soften somehow the short-distance core, for $|x|< \Lambda^{-1}$,
to make the Fourier transform possible. The resulting interaction at
momentum scales $\ll\Lambda$ will be dominated by ``contact'' terms
and will possess all the other unparticle properties discussed in
the introduction.

It is curious to note that the behavior of the propagator at both low
and high momenta is dictated by the short-distance part of the
interaction.  Indeed, we have established that the pure CFT
interaction at long distances contribute subdominantly to low-energy
scattering.
%The scattering amplitude for momenta $l^{-1}$ is
%controlled not by the behavior of the position-space propagator at
%distances $l$ (where it takes the pure CFT form), but by its
%short-distance regime (the ``contact'' terms). 
We already encountered this while examining the spectral
representation: the UV tail dominates the integral. Remarkably, ``long
distance physics'' and ``low energy physics'' are not the same.

\emph{Effective field theory view}. It is worth mentioning that the
last point in no way contradicts the general principles of effective
field theory. Indeed, regardless of the scale at which the different
parts of the interaction arise, at low energies the interaction is
described by a series of effective operators $\sim\Lambda^{-2} J^{SM}
J^{SM}$, $\sim\Lambda^{-4} J^{SM}
\partial^2 J^{SM}$, ..., plus the piece $\sim\Lambda^{-2d+2}
J^{SM}(p^2)^{(d-2)}J^{SM}$ that comes from the low-mass modes
that cannot be integrated out. For $d\ge3$ it then follows that the
contact terms coming from UV dominate in scattering. Notice that this
expansion should be directly compared to
Eqs.~(\ref{eq:spectralintegralcontact}) and
(\ref{eq:longitudinal_branetobrane_expand_Mink}). The latter, in
particular, fixes the relative coefficients of all the effective
operators, as should happen in any concrete model realizing the
unparticle scenario. 

\emph{On other realizations and generalizations}. In this paper, we
have considered a vector field in the bulk and shown that it possesses
all the properties of unparticles. The AdS/CFT conjecture is of course
known to be more general, and we believe that our results generalize
as well: bulk fields of other spins should also give unparticles of
corresponding spin ({\it cf.} \cite{Terning2008}). This RS
2/unparticle conjecture should be investigated further.

Other realizations of unparticles are certainly possible. Even within
the warped extra dimensions framework, the UV regime can be different,
as illustrated, {\it e.g.}, by the Lykken-Randall
\cite{lykken-randall} scenario treated in \cite{longpaper}. Depending
on the realization, the relative coefficients between the low-energy
effective operators will be different, but the general results
following from the effective field theory analysis must be preserved.
In particular, the dominance of the contact interactions should
persist, by simple power counting.

We close with the following observation. Phenomenologically, one may
be interested to consider a broader class of hidden sector models with
a continuous or quasi-continuous spectrum. Even though the underlying
theory may not be a strict CFT, it may turn out to look approximately
scale invariant in a range of energies accessible to a given
experiment ({\it cf.} QCD, \cite{Neubert}). In such a case, many of
the features of unparticles discussed here could still apply. This
particularly refers to the properties following from
Eq.~(\ref{eq:spectralintegralcontact}).  Whenever the spectral
integral is UV-dominated, the ``contact terms'' will dominate in SM-SM
scattering processes.  The unparticle phase space and the phase of the
CFT piece of the propagator are also derived from the spectral
representation and, so long as the latter looks ``sufficiently
conformal'' in the range of energies of the experiment, would follow
their unparticle forms. In this sense, they are more robust features
than the unitarity bounds, which, as we saw, could be avoided by
relaxing strict conformal symmetry, or considering gauge-variant operators.

\section{Conclusions}

In summary, at energies much below the AdS curvature, the RS 2 model
possesses the known properties of unparticle physics. This includes
both the CFT features -- the unitarity bounds and the tensor structure
-- and the features originating from the breaking of the CFT in the
UV, particularly the dominant ``contact'' terms.  We explicitly see,
as observed in \cite{GeorgiKats}, that the unparticle physics scenario
is not only about a scale invariant theory -- the CFT breaking and
coupling to the SM in the UV are its essential ingredients.

The UV regularization of the CFT by the flat five-dimensional space is
different from the Banks-Zaks scenario envisioned in the original work
\cite{GeorgiI}, nevertheless it is sufficient to control the
divergences of the spectral function and guarantee cancellations
between the ``contact'' terms and the CFT parts. At the same time, in
the infrared, the CFT is preserved, ensuring that the CFT tensor
structure and unitarity bounds are preserved.

The utility of having a concrete, treatable realization of the
unparticle scenario is demonstrated by the ease with which the known
properties of unparticles are obtained here, and, moreover, extended to
strong coupling. We hope that further applications of RS-like models
can shine light on other theoretical issues of unparticle physics.
Additionally, phenomenology of unparticles continues to be an active
area of research (see, {\it} e.g., \cite{Cheung2008} for a list of
references). It is hoped that having the RS 2 realization will lead to
further progress in this area as well.

\section*{Acknowledgments}

It is a pleasure to acknowledge conversations with T. Bhattacharya, L.
Randall, Yu.  Shirman, S. Thomas, and L. Vecchi. Preliminary results
of this work were presented at seminars at UC Berkeley and Caltech in
May 2008 and in more complete form in November 2008 at the Brookhaven
Forum 2008.  This work was supported by the U.S. Department of Energy
at Los Alamos National Laboratory under Contract No.
DE-AC52-06NA25396.

\end{document}